\documentclass[12pt]{article}

\usepackage{arxiv}

\usepackage[utf8]{inputenc} 
\usepackage[T1]{fontenc}    
\usepackage{hyperref}       
\usepackage{url}            
\usepackage{booktabs}       
\usepackage{amsfonts}       
\usepackage{nicefrac}       
\usepackage{microtype}      
\usepackage{lipsum}

\usepackage{amsmath}
\usepackage{multirow}
\usepackage{graphicx}
\usepackage{rotating}
\usepackage{array}
\usepackage{natbib}
\usepackage{amssymb}
\usepackage{float}

\title{Tuning Free Rank-Sparse Bayesian Matrix and Tensor Completion with Global-Local Priors}

\author{
  Daniel E. Gilbert \\
  Department of Statistics and Data Science\\
  Cornell University\\
  \texttt{deg257@cornell.edu} \\
   \And
  Martin T. Wells \\
  Department of Statistics and Data Science\\
  Cornell University\\
  \texttt{mtw1@cornell.edu} \\
}

\begin{document}
\maketitle

\begin{abstract}
Matrix and tensor completion are frameworks for a wide range of problems, including collaborative filtering, missing data, and image reconstruction. Missing entries are estimated by leveraging an assumption that the matrix or tensor is low-rank. Most existing Bayesian techniques encourage rank-sparsity by modelling factorized matrices and tensors with Normal-Gamma priors. However, the Horseshoe prior and other ``global-local'' formulations provide tuning-parameter-free solutions which may better achieve simultaneous rank-sparsity and missing-value recovery. We find these global-local priors outperform commonly used alternatives in simulations and in a collaborative filtering task predicting board game ratings.
\end{abstract}


\section{Introduction}

The problem of imputing missing values into possibly noisy matrices and tensors has wide-ranging applications. In missing data problems, practitioners use imputation to overcome the hurdle of non-response in service of building statistical models \citep{rubin2004multiple,sengupta2017,sportisse2018imputation}. In machine learning, various problems can be cast as missing data problems including collaborative filtering \citep{keshavan2010matrix,su2009survey}, as well as image and video denoising, reconstruction, and classification \citep{yuan2018higher,ji2010robust}. 

Imputation requires some assumption about the relationship between the missing values and the values which are observed. By assuming that the matrix or tensor is low-rank, we can specify the missing values using only a small number of observations. We will begin by discussing matrices, and later generalize the discussion to tensors.  

If the rank of the matrix is known, completing the matrix is a more straightforward problem \citep{geweke1996bayesian}. However, if the rank is unknown, simultaneously finding a suitable rank and completing the matrix is a non-convex and computationally complex problem \citep{candes2010matrix, candes2009exact}. A predominant frequentist approach yields a convex optimization problem by relaxing the low-rank assumption into a constraint on the nuclear norm of the matrix \citep{recht2011simpler,candes2010matrix}.

Meanwhile Bayesian methods use prior distributional information to encourage approximate rank sparsity in the matrix \textit{a posteriori}. Let $\mathbf{\Theta}$ be an $m \times n$ rank $r$ matrix with missing entries. $\mathbf{\Theta}$ can be decomposed into ``component'' matrices, $\mathbf{M}$ ($m \times K$) and $\mathbf{N}$ ($n \times K$) where $K \geq r$, such that $\mathbf{\Theta} = \mathbf{M}\mathbf{N}^\top = \sum_{k=1}^K \mathbf{M}_{\cdot k} \mathbf{N}_{\cdot k}^\top$. The Bayesian approach to adaptive rank estimation is to set $K > r$ and place priors distributions on $\mathbf{M}$ and $\mathbf{N}$ which encourage posterior estimates of entire columns to be exactly or approximately $0$.

The most common prior formulation \citep{babacan2011low,alquier2014bayesian,song2016bayesian} used to induce near-sparsity in the columns of the component matrices is the normal-gamma prior, here defined as:

\begin{equation*}
    \mathbf{M}_{\cdot k} \sim \mathcal{N}_m(0,\boldsymbol{\gamma}_k \mathbf{I}_m) \; \; \text{and} \; \mathbf{N}_{\cdot k} \sim \mathcal{N}_n(0,\boldsymbol{\gamma}_k \mathbf{I}_n)
\end{equation*}

where $\boldsymbol{\gamma}_k \overset{iid}{\sim} \mathcal{G}(\alpha,\beta)$ for $k \in \{1,\dots,K\}$. Here $\mathcal{G}(\cdot,\cdot)$ is the gamma distribution parametrized by shape and rate, respectively. By setting $\beta$ to be large, we can put posterior weight near $0$ for each $\gamma$ encouraging sparse solutions. 

For a solution which is unique up a factor of $-1$ on the columns of the component matrices, one can control the length and orthogonality of the columns of the component matrices by restricting them to Stiefel manifolds \citep{hoff2016equivariant,hoff2007model}. However, it is computationally expensive to do so, because it introduces complex dependencies \textit{a posteriori} between the all of the entries of the component matrices. By contrast, the unconstrained normal-gamma formulation does not specify unique posterior component matrices, but affords highly efficient block Gibbs sampling due to posterior row-wise independence in the component matrices. Uniquely specified component matrices are not always necessary as long as $\mathbf{\Theta}$ is estimated faithfully, and length and orthogonality constraints do not seem to bolster the estimation of $\mathbf{\Theta}$ in practice \citep{song2016bayesian,alquier2015bayesian}.  

This problem of selecting a small number of columns to remain non-null is similar to the problem of variable selection in linear regression. In that setting, choosing the best subset of variables to include in a model is computationally complex. A classic solution is the Lasso method, in which we relax the subset-cardinality constraint into an $\ell_1$ constraint \citep{tibshirani1996regression}. A Bayesian analogue to the Lasso is the application of the Laplace prior to model parameters \citep{park2008bayesian}. In this model, the \textit{maximum-a-posteriori} (MAP) parameter estimates coincide with the estimates produced by the Lasso.  

However, newly developed prior distributions have proven more effective in sparse regression than the Laplace prior, which fails to simultaneously induce sparsity while efficiently recovering non-null parameters \citep{van2016conditions}. Among these are ``global-local priors'' including the Horseshoe prior \citep{carvalho2010horseshoe}, the Horseshoe+ prior \citep{bhadra2017horseshoe+}, and most recently the Inverse-Gamma Gamma (IGG)  prior \citep{bai2017inverse}. 

Returning to the matrix completion problem, we see there is a connection between the normal-gamma prior and the Lasso. The Laplace distribution is an exponential scale-mixture of Gaussians, which is a special case of the normal-gamma prior. Indeed, MAP estimation in the normal-gamma case is equivalent to using a group-Lasso penalty on the columns of the component matrices \citep{alquier2014bayesian}. Having developed an instinctive aversion to all things Lasso and $\ell_1$-penalized, we will instead adapt these global-local priors to the group-case and use them to model the columns of the component matrices $\mathbf{M}$ and $\mathbf{N}$.

One advantage to these global-local priors is that they are tuning-parameter free. As we will demonstrate, the results using the normal-gamma prior are quite sensitive to the choice of $\beta$, which corresponds to the tuning parameter in the Lasso. This parameter ought to specify the level of sparsity, which is usually unknown \textit{a priori}. Learning the global shrinkage parameter with these global local priors avoids the need to use cross validation. Furthermore, the global-local priors result in sparser solutions and better estimates than the normal-gamma prior, both in simulations of low-rank incomplete matrices and in real data.

In this paper, We use the following notational conventions unless otherwise noted. Vectors are indicated by bolded lowercase letters and symbols, such as $\boldsymbol{\rho}$ and $\boldsymbol{\phi}$. Matrices and tensors are indicated by bolded uppercase letters and symbols such as $\mathbf{M}$ and $\mathbf{\Theta}$. $\mathbf{M}_{i \cdot}$ refers to a column vector formed by the entries of the $i$th row of the matrix $\mathbf{M}$, and $\mathbf{M}_{\cdot j}$ refers to a column vector formed by the entries of the $j$th column of the matrix $\mathbf{M}$. $\text{Vec}(\mathcal{A})$ refers to the column vector formed using the entries of a set of scalars, $\mathcal{A}$. For a length-n vector of indices $\mathbf{v}$, $\mathbf{M}_{\mathbf{v} j}$ is $\text{vec}(\{M_{\mathbf{v}_i j} \mid i \in \{1,\dots,n\} \}).$ Furthermore, $\mathbf{M}_{\mathbf{v} \cdot}$ refers to the matrix with $j$th column $\mathbf{M}_{\mathbf{v} j}$, and $\mathbf{M}_{i \cdot}$ refers to the matrix with $i$th row $\mathbf{M}_{i \mathbf{v}}$.

For two column vectors $\mathbf{v}$ and $\mathbf{w}$ length m and n respectively, $v \otimes w$ is their outer product, $\mathbf{v}\mathbf{w}^\top$. By extension, for several vectors $\mathbf{v}^{(1)}, \mathbf{v}^{(2)},\dots,\mathbf{v}^{(K)}$ with lengths $m_1,m_2,\dots,m_K$ respectively, $\mathbf{v}^{(1)} \otimes \mathbf{v}^{(2)} \otimes \dots \otimes \mathbf{v}^{(K)}$ is the $m_1 \times m_2 \times \dots \times m_K$ mode-K tensor $\mathbf{T}$ such that 

$$\mathbf{T}_{i_1,i_2,\dots,i_K} = v^{(1)}_{i_1} v^{(2)}_{i_2}\cdots v^{(K)}_{i_K}.$$ 

For two matrices of the same dimensions, $\mathbf{M}_1$ and $\mathbf{M}_2$, $\mathbf{M}_1 \odot \mathbf{M}_2$ indicates an element-wise matrix product.  

We use $\mathcal{N}(\cdot,\cdot)$ to denote the univariate normal distribution parametrized by mean and variance, and $\mathcal{N}_K(\cdot,\cdot)$ to denote the K-dimensional multivariate normal distribution parametrized by mean and dispersion. $C^+(\cdot,\cdot)$ is the Cauchy distribution truncated to the positive part, parametrized by location and scale. $\mathcal{G}(\cdot,\cdot)$ is the gamma distribution parametrized by shape and rate. $\mathcal{G}^{-1}(\cdot,\cdot)$ is the inverse gamma distribution parametrized by shape and scale. $\text{IG}(\cdot,\cdot)$ is the inverse Gaussian distribution parametrized by mean and shape, and $\text{GIG}(\cdot,\cdot,\cdot)$ is the generalized inverse Gaussian distribution parametrized as in \citet{johnson1994continuous}.

\section{Rank-Sparse Matrix Completion with Global-Local Priors}

Let $\mathbf{Y}$ be an $m \times n$ random matrix with a potentially large number of missing entries. Le $\mathcal{S} = \{(i,j)|Y_{ij}\text{ is not missing}\}.$ We will complete $\mathbf{Y}$ by assuming that it is  the sum of a centered mean matrix, row and column intercepts, an overall intercept, and Gaussian noise, that is,
    
\begin{equation} \label{Ymodel}
\mathbf{Y}_{ij} = \mathbf{\Theta_{ij}} + \boldsymbol{\rho}_i + \boldsymbol{\omega}_j + \mu + \mathbf{E}_{ij}
\end{equation}

where $\mathbf{\Theta}$ is an $m \times n$ mean matrix, $\boldsymbol{\rho}$ and $\boldsymbol{\omega}$ are vectors with lengths $m$ and $n$ respectively such that $\sum_{i=1}^m \boldsymbol{\rho}_i = 0$ and $\sum_{i=1}^n \boldsymbol{\omega}_i = 0$, $\mu$ is a scalar, and $\mathbf{E}$ is an $m \times n$ error matrix with $ \mathbf{E}_{ij} \overset{\text{i.i.d.}}{\sim} \mathcal{N}(0,\sigma^2).$

We also assume, the rank of $\mathbf{\Theta}$, $\text{rank}(\mathbf{\Theta}) \equiv r < \min(m,n)$, thus there exist (non-unique) $m \times r$ and $n \times r$ component matrices  $\mathbf{M^*}$ and $\mathbf{N^*}$ such that 
\begin{equation}\label{thetadecomp}
\mathbf{\Theta} = \mathbf{M^*}\mathbf{N^*}^\top.
\end{equation}

\subsection{Prior Formulation}

Under the  model in (\ref{Ymodel}) we aim to learn the true rank $r$ of $\mathbf{\Theta}.$ Expressing \ref{thetadecomp} in terms of vectors it follows that

$$\mathbf{\Theta} = \mathbf{M}\mathbf{N}^\top = \sum_{k=1}^K \mathbf{M}_{\cdot k}\mathbf{N}_{\cdot k}^\top$$
where $\mathbf{M}$ and $\mathbf{N}$ are $m \times K$ and $n \times K$ augmented component matrices, with $K > r$ columns. We choose $K > r$ so that we can assign continuous shrinkage priors to the columns of $\mathbf{M}$ and $\mathbf{N}$ to encourage all but $r$ columns to become small.
We formulate the mixture priors as

$$\mathbf{M}_{\cdot k}|\boldsymbol{\gamma}_k,\sigma \sim \mathcal{N}(\mathbf{0},\boldsymbol{\gamma}_k\sigma^2 \mathbf{I}_m ) \; \; \text{and} \;\; \mathbf{N}_{\cdot k}|\boldsymbol{\boldsymbol{\gamma}_k},\sigma \sim \mathcal{N}(\mathbf{0},\boldsymbol{\gamma}_k\sigma^2 \mathbf{I}_n) \text{\hspace{5mm} for $k = 1,\dots,K$},    $$

\;

$\sigma^2 \sim \mathcal{G}^{-1}(a_\sigma, b_\sigma)$, and to the the row and column intercept vectors $\rho$ and $\omega$ in (\ref{Ymodel}) we assign independent improper uniform priors.

We can express all of the prior formulations we will use in this paper in terms of the distribution of $\gamma_k$ for $k = 1,\dots,K$ as well as some fixed hyperparameters $\alpha$, $\beta$, $a$, $b$, and $c$. The standard priors are taken from \citet{alquier2014bayesian}. Following \citet{alquier2014bayesian}, we set shape parameter for the gamma prior to $\alpha = (m + n + 1)/2$.

\begin{table}[H]
\centering
\caption{Prior Formulations for $\mathbf{\gamma}$} 
\vspace{3mm}
\label{prior_formulations}

\begin{tabular}{c | c l c l}
& Prior & \multicolumn{3}{c}{Column Variance Distribution} \\ \hline
\\
\multirow{3}{*}{Traditional} & Gaussian & $\boldsymbol{\gamma}_k = V_0$ (constant)   \\ 
\\
& Gamma & $\boldsymbol{\gamma}_k \overset{\text{i.i.d.}}{\sim} \mathcal{G}(\alpha,\beta)$ \\
\\
\hline
\\
\multirow{8}{*}{Global-Local} & Horseshoe & $\boldsymbol{\gamma}_k \overset{\text{D}}{=}  \lambda_k^2 \tau^2$ & where &
$\lambda_k^2 \overset{\text{i.i.d.}}{\sim} C^+(0,1)$ \\
& & & & $\tau^2 \sim C^+(0,1)$ \\ 
\\
& Horseshoe+ & $\boldsymbol{\gamma}_k \overset{\text{D}}{=} \lambda_k^2 \eta_k^2 \tau^2$ & where &
$\lambda_k^2 \overset{\text{i.i.d.}}{\sim} C^+(0,1)$ \\
& & & & $\eta_k^2 \overset{\text{i.i.d.}}{\sim} C^+(0,1)$ \\
& & & & $\tau^2 \sim C^+(0,1)$ \\ 
\\
& GIG & $\boldsymbol{\gamma}_k \overset{\text{D}}{=}  \lambda_k \tau_k$ & where &
$\lambda_k \overset{\text{i.i.d.}}{\sim} \mathcal{G}^{-1}(a,c)$ \\
& & & & $\tau_k \overset{\text{i.i.d.}}{\sim} \mathcal{G}(b,c)$  
\end{tabular}

\end{table}

If the performance of a method is highly sensitive to the prior hyperparameters, these hyperparameters must be treated as tuning parameters. In the unlikely situation we know the true level of rank sparsity of the matrix to be completed, we can choose hyperparameters using this information. Otherwise, we will need to use a sample-data-dependent technique such as cross-validation. We find in simulations that the Gaussian and Gamma methods are sensitive to the choice of $V_0$ and $\beta$, respectively.

However, the Horseshoe and Horseshoe+ methods contain no prior hyperparameters, as the hierarchical prior structure allows the mixture components of $\boldsymbol{\gamma}$ to adapt to the level of sparsity in the data. We therefore consider these methods ``tuning parameter free,'' and in practice practitioners can use these priors without first finding suitable hyperparameter values. The Gamma Inverse-Gamma prior has a similar property, as we will keep its prior hyperparameters constant. Due to its global-local construction, this posterior distributions corresponding to the Gamma Inverse-Gamma prior are not particularly sensitive to the choices of $a$, $b$, and $c$.

\subsection{Gibbs Sampler}

We explore the posterior distribution of $\mathbf{\Theta}$, $\sigma$, $\boldsymbol{\rho}$ and $\boldsymbol{\omega}$ using Gibbs sampling. Efficient block sampling is facilitated by the posterior independence of the rows of the component matrices, $\mathbf{M}$ and $\mathbf{N}$. 
Define the sets of indices
$$\mathbf{j}_i = \{j \in 1,\dots,n | (i,j) \in \mathcal S\} \text{ for $i = 1,\dots,m$} $$
$$\mathbf{i}_j = \{i \in 1,\dots,m| (i,j) \in \mathcal S \} \text{ for $j = 1,\dots,n$} $$
Thus $\mathbf{j}_i$ is a vector of column indices associated with non-missing entries in the $i$th row of $\mathbf{Y}$, and $\mathbf{i}_j$ is a vector of row indices associated with non-missing entries in the $j$th column of $\mathbf{Y}$.  Let $\mathbf{\Gamma}$ be the $K \times K$ diagonal matrix such that $\mathbf{\Gamma}_{kk} = \boldsymbol{\gamma}_k$.


The Gibbs sampling algorithm is run by iteratively sampling each parameter or parameter block from its ``full conditional distribution,'' meaning its distribution conditional on the data, $\{Y_{ij} \mid (i,j) \in \mathcal{S}\}$, as well as every other model parameter upon which it depends. For some model parameter $\theta$, the expression $\theta \mid \cdot \sim F$ indicates that $F$ is the full conditional distribution of $\theta$. The full conditional distributions for the component matrices are:

\begin{align*}
    \mathbf{M}_{i\cdot}|\cdot &\sim \mathcal{N}_K\left(\left(\mathbf{N}_{\mathbf{j}_i \cdot}^\top \mathbf{N}_{\mathbf{j}_i \cdot}
 + \mathbf{\Gamma}^{-1}\right)^{-1}\mathbf{N}_{\mathbf{j}_i \cdot}^\top \mathbf{Y}_{i \mathbf{j}_i}, \sigma^2 \left( \mathbf{N}_{\mathbf{j}_i \cdot}^\top \mathbf{N}_{\mathbf{j}_i \cdot}
 + \mathbf{\Gamma}^{-1}\right)^{-1}\right) \hfill &&\text{for } i \in 1,\dots,m\\
 \mathbf{N}_{j \cdot} | \cdot &\sim \mathcal{N}_K\left(\left(\mathbf{M}_{\cdot \mathbf{i}_j }^\top \mathbf{M}_{\cdot \mathbf{i}_j }
 + \mathbf{\Gamma}^{-1}\right)^{-1}\mathbf{M}_{\cdot \mathbf{i}_j }^\top \mathbf{Y}_{\mathbf{i}_j j}, \sigma^2 \left( \mathbf{M}_{\cdot \mathbf{i}_j }^\top \mathbf{M}_{\cdot \mathbf{i}_j }
 + \mathbf{\Gamma}^{-1}\right)^{-1}\right) \hfill &&\text{for } j \in 1,\dots,n,
\end{align*}

with independence between rows of each component matrix. Therefore the rows of the component matrices can be sampled as blocks, even in parallel.

The full conditional distributions for the intercepts are:

\begin{align*}
    \boldsymbol{\rho}_i \mid \cdot &\overset{\text{i.i.d.}}{\sim} \mathcal{N}\left( \frac{1}{|\mathbf{j}_i|}\sum_{j \in \mathbf{j}_i} (\mathbf{Y}_{ij} - \mathbf{\Theta}_{ij} - \boldsymbol{\omega}_j - \mu), \frac{\sigma^2}{|\mathbf{j}_i|}\right)\mathbf{1}_{\boldsymbol{\rho}^\top \mathbf{1} = 0} \text{ for } i \in 1,\dots,m \\
    \boldsymbol{\omega}_j \mid \cdot &\overset{\text{i.i.d.}}{\sim} \mathcal{N}\left( \frac{1}{|\mathbf{i}_j|}\sum_{i \in \mathbf{i}_j} (\mathbf{Y}_{ij} - \mathbf{\Theta}_{ij} - \boldsymbol{\rho}_i - \mu), \frac{\sigma^2}{|\mathbf{j}_i|}\right)\mathbf{1}_{\boldsymbol{\rho}^\top \mathbf{1} = 0} \text{ for } i \in 1,\dots,m \\
    \mu \mid \cdot &\sim \mathcal{N}\left(\frac{1}{|\mathcal{S}|}\sum_{\{i,j\} \in \mathcal{S}} (\mathbf{Y}_{ij} - \mathbf{\Theta}_{ij} - \boldsymbol{\rho}_i - \boldsymbol{\omega}_j), \frac{\sigma^2}{|\mathcal{S}|}\right).
\end{align*}

Note that for large matrices, sampling from the constrained posterior distributions for $\boldsymbol{\rho}$ and $\boldsymbol{\omega}$ is computationally expensive, but in practice unnecessary. Sampling from the unconstrained Gaussians and stabilizing the sampling routine by recentering these vectors after each iteration achieves nearly identical results. 

The full conditional distribution for the noise variance is:
\begin{align*}
    \sigma^2 \mid \cdot \sim \Gamma^{-1} \left( a_\sigma + \frac{|\mathcal{S}|}{2},b_\sigma + \frac{1}{2}\sum_{{i,j} \in \mathcal{S}} (\mathbf{Y}_{ij} - \mathbf{\Theta}_{ij} - \boldsymbol{\rho}_i - \boldsymbol{\omega}_j - \mu)^2\right).
\end{align*}

Finally, only the way in which we sample the column variances $\boldsymbol{\gamma}$ depends on our choice of prior formulation.

\subsubsection*{Gaussian Priors}

Here the column variances are constant. Thus $\boldsymbol{\gamma}_k \mid \cdot = V_0$ for all $k$ in $\{1,\dots,K\}$.  

\subsubsection*{Gamma Priors}

Here, the gamma prior is a special case of the Generalized Inverse Gaussian distribution, which is conjugate to the Gaussian component column entries \citep{johnson1994continuous}. Given the careful choice of $\alpha = \frac{m+n+1}{2}$ from \citet{alquier2014bayesian}, the posterior simplifies to a an Inverse Gaussian distributed random variable. Thus

$$\boldsymbol{\gamma}_k \mid \cdot \overset{\text{i.i.d.}}{\sim} \text{IG}\left(\frac{\beta}{||\mathbf{M}_{\cdot k}||_2^2 + ||\mathbf{N}_{\cdot k}||_2^2}, \beta^2 \right) \text{for } k \in 1,\dots,K.$$

\subsubsection*{Horseshoe Priors}

Following \citet{makalic2016simple}, we represent the positive-Cauchy-distributed variance terms as mixtures of inverse-gammas to facilitate sampling. For the Horseshoe formulation, \textit{a priori}:

    $$\lambda_k^2 \mid \nu_k \sim \mathcal{G}^{-1}(\frac{1}{2}, \frac{1}{\nu_k})
    \text{ and } \nu_k \sim \mathcal{G}^{-1}(\frac{1}{2}, 1) \hspace{5mm} \text{ for } k \in 1,\dots,K,$$ 
    $$    \tau^2 \mid \xi \sim \mathcal{G}^{-1}(\frac{1}{2},\frac{1}{\xi}) \text{ and }
    \xi \sim \mathcal{G}^{-1}(\frac{1}{2},1). $$

Thus \textit{a posteriori}, the complete conditional distributions are:

\begin{align*}
    \lambda^2_k \mid \cdot &\overset{\text{i.i.d.}}{\sim} \mathcal{G}^{-1}\left(\frac{1 + m + n}{2},\frac{1}{\nu_k} + \frac{||\mathbf{M}_{\cdot k}||_2^2 + ||\mathbf{N}_{\cdot k}||_2^2}{2\tau^2\sigma^2}\right) \hspace{5mm} \text{for } k \in 1,\dots,K\\
    \nu_k \mid \cdot &\overset{\text{i.i.d.}}{\sim} \mathcal{G}^{-1}\left(1,1+\frac{1}{\lambda_k^2} \right) \\
    \tau^2 \mid \cdot &\sim \mathcal{G}^{-1}\left(\frac{1 + K(m+n)}{2}, \frac{1}{\xi} + \sum_{k=1}^K \left(\frac{||\mathbf{M}_{\cdot k}||_2^2 + ||\mathbf{N}_{\cdot k}||_2^2}{2\lambda^2_k\sigma^2}\right)  \right) \\
    \xi \mid \cdot &\sim \mathcal{G}^{-1}\left(1,1+\frac{1}{\tau^2} \right).
\end{align*}

\subsubsection*{Horseshoe+ Priors}

For the Horseshoe+ prior we simply add in an extra pair of hierarchical variance terms, so we have

  $$\lambda_k^2 \mid \nu_k \sim \mathcal{G}^{-1}(\frac{1}{2}, \frac{1}{\nu_k})
    \text{ and } \nu_k \sim \mathcal{G}^{-1}(\frac{1}{2}, 1) \hspace{5mm} \text{ for } k \in 1,\dots,K,$$ 
      $$\eta_k^2 \mid \phi_k \sim \mathcal{G}^{-1}(\frac{1}{2}, \frac{1}{\phi_k})
    \text{ and } \nu_k \sim \mathcal{G}^{-1}(\frac{1}{2}, 1) \hspace{5mm} \text{ for } k \in 1,\dots,K,$$ 
    $$    \tau^2 \mid \xi \sim \mathcal{G}^{-1}(\frac{1}{2},\frac{1}{\xi}) \text{ and }
    \xi \sim \mathcal{G}^{-1}(\frac{1}{2},1). $$
 
Thus \textit{a posteriori}, the complete conditional distributions are:

\begin{align*}
    \lambda^2_k \mid \cdot &\overset{\text{i.i.d.}}{\sim} \mathcal{G}^{-1}\left(\frac{1 + m + n}{2},\frac{1}{\nu_k} + \frac{||\mathbf{M}_{\cdot k}||_2^2 + ||\mathbf{N}_{\cdot k}||_2^2}{2\eta_k^2\tau^2\sigma^2}\right) \hspace{5mm} \text{for } k \in 1,\dots,K\\
    \nu_k \mid \cdot &\overset{\text{i.i.d.}}{\sim} \mathcal{G}^{-1}\left(1,1+\frac{1}{\lambda_k^2} \right) \hspace{5mm} \text{for } k \in 1,\dots,K \\
        \eta^2_k \mid \cdot &\overset{\text{i.i.d.}}{\sim} \mathcal{G}^{-1}\left(\frac{1 + m + n}{2},\frac{1}{\phi_k} + \frac{||\mathbf{M}_{\cdot k}||_2^2 + ||\mathbf{N}_{\cdot k}||_2^2}{2\lambda_k^2\tau^2\sigma^2}\right) \hspace{5mm} \text{for } k \in 1,\dots,K\\
    \phi_k \mid \cdot &\overset{\text{i.i.d.}}{\sim} \mathcal{G}^{-1}\left(1,1+\frac{1}{\eta_k^2} \right) \hspace{5mm} \text{for } k \in 1,\dots,K \\
    \tau^2 \mid \cdot &\sim \mathcal{G}^{-1}\left(\frac{1 + K(m+n)}{2}, \frac{1}{\xi} + \sum_{k=1}^K \left(\frac{||\mathbf{M}_{\cdot k}||_2^2 + ||\mathbf{N}_{\cdot k}||_2^2}{2\lambda^2_k\eta^2_k\sigma^2}\right)  \right) \\
    \xi \mid \cdot &\sim \mathcal{G}^{-1}\left(1,1+\frac{1}{\tau^2} \right).
\end{align*}

\subsubsection*{Inverse-Gamma Gamma Priors}

For the IGG formulation, the gamma distribution again enjoys conjugacy to the component column as a special case of the generalized inverse Gaussian distribution, and the inverse gamma distribution is itself conjugate to the component column. Therefore, \textit{a posteriori}, the complete conditional distributions are:

\begin{align*}
    \tau_k \mid \cdot &\overset{\text{i.i.d.}}{\sim} \text{GIG}\left(2c, \frac{||\mathbf{M}_{\cdot h}||_2^2 + ||\mathbf{N}_{\cdot h}||_2^2}{\lambda_k\sigma^2}, b - \frac{m+n}{2}\right) \\
    \lambda_k \mid \cdot &\overset{\text{i.i.d.}}{\sim} \mathcal{G}^{-1}\left(a + \frac{m + n}{2}, c + \frac{||\mathbf{M}_{\cdot h}||_2^2 + ||\mathbf{N}_{\cdot h}||_2^2}{2\tau_k\sigma^2} \right).
\end{align*}

As in the Gamma formulation, if we set $b = \frac{m + n + 1}{2}$, the generalized inverse Gaussian distribution posterior reduces to an inverse Gaussian posterior. 

Using the representation of the positive-Cauchy distribution as an inverse-gamma mixture of inverse-gammas, we can see that all of these prior formulations are similar in that the component column variance terms $\boldsymbol{\gamma}$ are products of some number and combination of gamma and inverse-gamma distributed global (column-spanning) and local (column-specific) factors.

\section{Extension to Tensor Completion}

In the case of multiway data, we can easily extend this methodology to tensors. Many frequentist and Bayesian solutions rely on the CANDECOMP/PARAFAC or ``CP'' tensor decomposition \citep{kolda2009tensor, zhao2016bayesian,bazerque2013rank}. We notice that by extending the matrix formulation by adding additional component matrices, we achieve a CP-like decomposition, although, as in the matrix case, we will not enforce the orthogonality of the columns of the component matrices. 

Suppose that $\mathbf{Y} \in \Re^{m_1 \times m_2 \times \dots \times m_D}$ is a $D$-dimensional tensor with a large proportion of missing values. Now $\mathcal{S} = \{(i_1,i_2,\dots,i_D)|Y_{i_1 i_2 \dots i_D} \text{ is not missing}\}$ 

Again, suppose the tensor $\mathbf{Y}$ is observed with Gaussian noise, thus
\begin{equation}\label{tensorY}
\mathbf{Y} = \mathbf{\Theta} + \mathbf{E}
\end{equation}
    
where $\mathbf{\Theta}$ is an $ m_1 \times m_2 \times \dots \times m_D$ mean matrix and $\mathbf{E}$ is an $m_1 \times m_2 \times \dots \times m_D$ error matrix with $ \mathbf{E}_{i_1 i_2 \dots i_D} \overset{\text{iid}}{\sim} N(0,\sigma^2).$ For notational simplicity, we will assume $\mathbf{\Theta}$ is centered and forego intercept terms in (\ref{tensorY}). However, the incorporation of intercepts corresponding to any dimensions is straightforward and analogous to the matrix case in \ref{Ymodel}. 
    
The rank of $\mathbf{\Theta}$, $\text{rank}(\mathbf{\Theta}) \equiv r < \min(m_1,m_2,\dots,m_D)$, thus there exist (non-unique) component matrices $\left\{\mathbf{M}^{(d)}, d \in 1,\dots,D \right\}$ with dimensions $m_d \times r$ such that

\begin{equation}\label{tensordecomp}
\mathbf{\Theta} = \sum_{k = 1}^r \mathbf{M}^{(1)}_{\cdot k} \otimes \mathbf{M}^{(2)}_{\cdot k} \otimes \dots \otimes \mathbf{M}^{(D)}_{\cdot k}
\end{equation}

Again, we aim to learn the true rank $r$ of $\mathbf{\Theta}$ in (\ref{tensordecomp}). We will augment the component matrices to $K>r$ rows such that

 $$\mathbf{\Theta} = \sum_{k = 1}^K \mathbf{M}^{(1)}_{\cdot k} \otimes \mathbf{M}^{(2)}_{\cdot k} \otimes \dots \otimes \mathbf{M}^{(D)}_{\cdot k}$$
where $\left\{\mathbf{M}^{(d)}, d \in 1,\dots,D \right\}$ have dimensions $m_d \times K$. Now

$$\mathbf{M}^{(d)}_{\cdot k}|\boldsymbol{\gamma}_k,\sigma \sim \mathcal{N}(\mathbf{0},\boldsymbol{\gamma}_k\sigma^2 \mathbf{I}_m ) \text{\hspace{5mm} for $k = 1,\dots,K$ \hspace{2mm} $d = 1,\dots,D$} $$
independent across columns and component matrices. The prior formulations for $\boldsymbol{\gamma}$ are identical to the matrix case; see to Table \ref{prior_formulations}. As before, 
$$\sigma^2 \sim \mathcal{G}^{-1}(a_\sigma,b_\sigma).$$

\subsection{Gibbs Sampler}

Again, efficient block sampling is facilitated by the posterior independence of the rows of the component matrices, $\mathbf{M}^{(d)}, d \in 1,\dots,d$.
Define the indices

$$\mathbf{i}^{(d_1,d_2)}_l = \{i \in 1,\dots,m_{d_2} | \text{ } \exists \text{ } (i_1,\dots,i_D) \in \mathcal{S} \text{ s.t. } i_{d_1} = l, i_{d_2} = i\} \text{ for $l = 1,\dots,m_{d_1}$}. $$

Thus $\mathbf{i}^{(d_1,d_2)}_l$ is a vector of indices along dimension $d_2$ corresponding to the set of entries where $i_{d_1} = l$.
Now let

$$\mathbf{N}^{(d_1)}_{(l)} = \overset{D}{\underset{\substack{d_2 = 1 \\ d_2 \neq d_1}}{\odot}} \mathbf{M}^{(d_2)}_{\mathbf{i}^{(d_1,d_2)}_l \cdot},$$
where $\odot$ is an elementwise matrix product.
Let $\mathbf{Y}^{(d)}_{(l)} = \text{vec}\{Y_{i_1,\dots,i_D}|i_d = l\}$. Again, let $\mathbf{\Gamma}$ be the $K \times K$ diagonal matrix such that $\mathbf{\Gamma}_{kk} = \boldsymbol{\gamma}_k$.


Thus, the full conditional distributions can be expressed as
$$\mathbf{M}^{d}_{l\cdot} \mid \cdot \sim \mathcal{N}_K\left(\left({\mathbf{N}^{(d)}_{(l)}}^\top \mathbf{N}^{(d)}_{(l)} + \mathbf{\Gamma}^{-1}\right)^{-1} {\mathbf{N}^{(d)}_{(l)}}^\top Y^{(d)}_{(l)}, \sigma^2 \left( {\mathbf{N}^{(d)}_{(l)}}^\top \mathbf{N}^{(d)}_{(l)} + \mathbf{\Gamma}^{-1}\right)^{-1}\right).$$

for $d \in 1,\dots,D$ and $l \in 1,\dots,m_d$. Furthermore,

$$\sigma^2|\cdot \sim \Gamma^{-1}\left(a_\sigma + \frac{|\mathcal{S}|}{2}, b_\sigma + \frac{1}{2|\mathcal{S}|}\sum_{(i_1,\dots,i_D) \in \mathcal{S}} \left( \mathbf{Y}_{i_1,\dots,i_D} - \left( \sum_{k=1}^K M^{(1)}_{i_1 k} \otimes M^{(2)}_{i_2 k} \otimes \dots \otimes M^{(D)}_{i_D k} \right) \right)^2\right)$$

All of the posterior sampling distributions from the component column variances $\mathbf{\gamma}$ can be preserved from the matrix case with only slight modification. Those terms which depend on columns of the component matrices $\mathbf{M}$ and $\mathbf{N}$ in the matrix case must be extended to depend on the corresponding columns from all of the component matrices $\mathbf{M}^{(1)},\dots,\mathbf{M}^{(D)}$ in the tensor case. For example, the full conditional distributions in the case of the horseshoe formulation are: 

\begin{align*}
    \lambda_k^2 \mid\cdot &\overset{\text{i.i.d.}}{\sim} \mathcal{G}^{-1}\left(\frac{1 + \sum_{d=1}^D m_d}{2},\frac{1}{\nu_k} + \frac{\sum_{d=1}^D||\mathbf{M^{(d)}}_{\cdot k}||_2^2}{2\tau^2\sigma^2}\right) \hspace{5mm} \text{for } k \in 1,\dots,K\\
    \nu_k \mid \cdot &\overset{\text{i.i.d.}}{\sim} \mathcal{G}^{-1}\left(1,1+\frac{1}{\lambda_k^2}\right)  \text{for } k \in 1,\dots,K \\
    \tau^2\mid\cdot &\sim \mathcal{G}^{-1}\left(\frac{1 + K(\sum_{d=1}^D m_d)}{2}, \frac{1}{\xi} + \sum_{k=1}^K \left(\frac{\sum_{d=1}^D||\mathbf{M^{(d)}}_{\cdot k}||_2^2}{2\lambda^2_k\sigma^2}\right)  \right) \\
    \xi\mid \cdot &\sim \mathcal{G}^{-1}\left(1,1+\frac{1}{\tau^2}\right). 
\end{align*}

\section{Simulation Studies}

In the following simulation studies, we compare the performance of the proposed global-local formulations to the Gaussian and Gamma priors. In the 2-dimensional (matrix) case, we also compare our results to a popular frequentist matrix completion algorithm, softImpute \citep{hastie2015matrix}, which is based on nuclear norm regularization. In each simulation, we generate random low-rank matrices and tensors using the following procedure. Given the desired dimensions $m_1,\dots,m_D$, true rank $r$ and scalar column variances $v_l$, $l \in \{1,\dots,r\}$, we generate $D$ component matrices in which each entry $M^{(d)}_{ij}$ is an independent draw from $\text{N}(0,v_l)$. The simulated mean matrix or tensor is then $$\mathbf{\Theta} = \sum_{k = 1}^r \mathbf{M}^{(1)}_{\cdot k} \otimes \mathbf{M}^{(2)}_{\cdot k} \otimes \dots \otimes \mathbf{M}^{(D)}_{\cdot k}$$ to which we add Gaussian noise, thus

$$\mathbf{Y} = \mathbf{\Theta} + \mathbf{E}$$
where $$\mathbf{E}_{i_1 i_2 \dots i_D} \overset{\text{i.i.d.}}{\sim} \mathcal{N}(0,\sigma^2).$$

We then select a proportion $p$ of entries to retain uniformly at random, with the condition that at least one entry from each row and column (and slice, etc.) remain.

The softImpute algorithm requires a tuning parameter, $\lambda$, which specifies the desired level of rank-sparsity. The Gamma formulation requires a prior hyperparameter, $\beta$, which serves much the same function. In these simulations, we will be unusually charitable towards these algorithms by training their tuning parameters using oracle information for each combination of test parameters. 

The Gaussian and IGG formulations also require prior hyperparameters, but we find ther results of these algorithms are not particularly sensitive to the choice of hyperparameters. For the Gaussian case, it is necessarily only to choose $V_0$ to be sufficiently large. In the IGG case, following \citet{bai2017inverse}, we set $a=1$, $c=1$, and $b<.5$.  

\subsection{Changing the rank-sparsity} 

Here we study the performance of the various algorithms under changing levels of rank-sparsity.

\subsubsection{Matrix Case}

Here we simulate $100 \times 100$ matrices, setting $\sigma^2 = .5$ and $v_l = 5$ for $l \in \{1,\dots,r\}$ and keeping $p=20\%$ of observations. 

For the Bayesian algorithms, we set $K=20$. We also set the max rank setting in softImpute to $20$. We attain Bayes estimates from the Gibbs samplers using 100 samples with a thinning factor of 5 after a burn-in of 500 iterations. 

For a given estimate of $\mathbf{\Theta}$, denoted $\mathbf{\hat{\Theta}},$ We calculate the standard error as:

$$\text{SE} = \sqrt{\sum_{i=1}^m\sum_{j=1}^n (\mathbf{\hat\Theta}_{ij} - \mathbf{\Theta}_{ij})^2}.$$

The standard errors presented are averaged over 100 trials.

\begin{table}[H]
    \centering
\begin{tabular}{r | l l l l l}
     &   & \multicolumn{4}{c}{$r$} \\
     &   & 2 & 4 & 8 & 16 \\ \hline
Horseshoe+ & {SE} & \bf{.374} & \bf{.608} & \bf{1.50} & \bf{15.8} \\ \hline
Horseshoe & {SE} & {.375} & {.608} & {1.50} & {15.8} \\ \hline
\multirow{4}{*}{IGG}    & a &  1 & \multicolumn{3}{c}{$\xrightarrow{\hspace*{2.5cm}}$} \\
    & b & .4 & \multicolumn{3}{c}{$\xrightarrow{\hspace*{2.5cm}}$} \\ 
    & c &  1 & \multicolumn{3}{c}{$\xrightarrow{\hspace*{2.5cm}}$} \\
    & {SE} & {.397} & {.624} & {1.55} & {16.7} \\ \hline
\multirow{2}{*}{Gamma} & $\beta$ & 40 & 27 & 12 & 10 \\
    & {SE} &  {.385} & {.631} & {1.95} & {16.2} \\ \hline
\multirow{2}{*}{Gaussian} & $V_0$ & 10 & \multicolumn{3}{c}{$\xrightarrow{\hspace*{2.5cm}}$} \\
    & {SE} & {.654} & {1.29} & {7.17} & {17.1} \\ \hline
\multirow{2}{*}{softImpute} & $\lambda$ & 10.4 & 12.0 & 18.4 & 28.2 \\
    & {SE} &  {1.50} & {3.29} & {8.50} & {15.9}  
\end{tabular}
    \caption{Chosen hyperparameters and performance of several algorithms over various degrees of rank-sparsity.}
    \label{2dvarr}
\end{table}

We find that even when competing algorithms are tuned with oracle information, the global-local priors still achieve the lowest SE, especially at high levels of sparsity. All of the Bayesian methods outperform softImpute at high levels of sparsity, but in high-dimensional cases where there is very little information about the missing entries, softImpute is competitive. 

\subsubsection{Order-3 Tensor Case}

We find similar results for order-3 tensors. The tensors we generate have $\sigma^2 = .5$, $v_l = 5$ for $l \in \{1,\dots,r\}$ and dimensions $20 \times 20 \times 25$, keeping only $p=10\%$ of the entries. For the Bayesian algorithms, we use the same settings as above and tune the gamma prior using oracle information. 

Similar to above, the standard error is:

$$\text{SE} = ||\text{vec}(\mathbf{\hat{\Theta}}) - \text{vec}(\mathbf{\Theta}) ||_2.$$

The standard errors presented are averaged over 100 iterations.

\begin{table}[H]
    \centering
\begin{tabular}{r | l l l l l}
     &   & \multicolumn{4}{c}{$r$} \\
     &   & 2 & 4 & 8 & 16 \\ \hline
Horseshoe+ & {SE} & \bf{.305} & {.463} & \bf{.828} & \bf{35.8} \\ \hline
Horseshoe & {SE} & {.306} & \bf{.461} & \bf{.828} & {36.7} \\ \hline
\multirow{4}{*}{IGG}    & a &  1 & \multicolumn{3}{c}{$\xrightarrow{\hspace*{2.5cm}}$} \\
    & b & .4 & \multicolumn{3}{c}{$\xrightarrow{\hspace*{2.5cm}}$} \\ 
    & c &  1 & \multicolumn{3}{c}{$\xrightarrow{\hspace*{2.5cm}}$} \\
    & {SE} & {.318} & {.471} & {.835} & {42.7} \\ \hline
\multirow{2}{*}{Gamma} & $\beta$ & 20 & 17 & 15 & 15 \\
    & {SE} &  \bf{.306} & \bf{.461} & {.835} & {42.7} \\ \hline
\multirow{2}{*}{Gaussian} & $V_0$ & 10 & \multicolumn{3}{c}{$\xrightarrow{\hspace*{2.5cm}}$} \\
    & {SE} & {.425} & {.568} & {21.2} & {62.2}
\end{tabular}
    \caption{Chosen hyperparameters and performance of several algorithms over various degrees of rank-sparsity.}
    \label{3dvarr}
\end{table}

Again, we find that even when the gamma algorithm is tuned with oracle information, the global-local priors still achieve the same or lower SE, especially at high levels of sparsity.

\subsection{Changing the proportion of missing entries}

In this experiment, we vary the level of missing observations in $100 \times 100$ matrices while keeping the true rank fixed at $r = 4$ with column variances $v_l = 5$ for $l \in \{1,\dots,4\}$. As above, for the Bayesian algorithms, we set $K=20$. We also set the max rank setting in softImpute to $20$. We attain Bayes estimates from the Gibbs samplers using 100 samples with a thinning factor of 5 after a burn-in of 500 iterations. Again, the standard errors are averaged over 100 iterations.

\begin{table}[H]
    \centering
    \begin{tabular}{r | l l l l l l}
     &   & \multicolumn{5}{c}{$p$} \\
     &   & .075 & .15 & .3 & .8 & 1 \\ \hline
Horseshoe+ & {SE} & {6.39} & \bf{.860} & \bf{.425} & \bf{.226} & \bf{.200} \\ \hline
Horseshoe & {SE} & \bf{6.36} & \bf{.860} & \bf{.425} & {.227} & {.201}\\ \hline
\multirow{4}{*}{IGG}    & a &  1 & \multicolumn{4}{c}{$\xrightarrow{\hspace*{3.5cm}}$}\\
    & b &  .4 & \multicolumn{4}{c}{$\xrightarrow{\hspace*{3.5cm}}$}\\ 
    & c &  1 & \multicolumn{4}{c}{$\xrightarrow{\hspace*{3.5cm}}$} \\
    & {SE} & {6.68} & {.878} & {.440} & {.240} & {.213} \\ \hline
\multirow{2}{*}{Gamma} & $\beta$ & 15 & 30 & 40 & 50 & 60 \\
    & {SE} &  {13.7} & {1.03} & {.438} & {.229} & {.201} \\ \hline
\multirow{2}{*}{Gaussian} & $V_0$ & 10 & \multicolumn{4}{c}{$\xrightarrow{\hspace*{3.5cm}}$} \\
    & {SE} & {13.1} & {5.37} & {.682} & {.424} & {.366} \\ \hline
\multirow{2}{*}{SoftImpute} & $\lambda$ & 15 & 13 & 8 & 8 & 8 \\
    & {SE} &  {8.57} & {5.13} & {1.62} & {.349} & {.366}  
\end{tabular}
    \caption{Chosen hyperparameters and performance of several algorithms over varying levels of missing data.}
    \label{2dvarp}
\end{table}

Again, the Horseshoe+ prior formulation generally achieves the best performance, although it is very similar to the performance of the Horseshoe. The Gamma and Gaussian priors perform particularly poorly when there are extremely few observations. The sparsity inducing structures in the global-local prior formulations seem essential in rank-sparse scenarios when there are too few observations for the standard Gaussian formulation to approximate the column variances using the observed data. Here, the performance of the softImpute algorithm largely mimics the performance of the Gaussian algorithm.

\subsection{Rank-Sparsity Recovery}

Here we showcase the tendency of the global-local Bayesian algorithms to faithfully identify the true underlying dimensionality of the matrix to be completed. We simulate $100 \times 100$ matrices with rank $r=6$ and set column variances $\mathbf{v} = \{6,6,3,3,1\}.$ Here we retain only $p = .15\%$ of the observations. This scree plot (with values averaged over 100 iterations) demonstrates the the fact that the output of the Bayesian algorithms concentrates more of the information about the signal in the first five singular vectors. The Gaussian algorithm and softImpute disperse more of the information over the full allotment of dimensions. Thus in rank-sparse cases, these sparser formulations may help more accurately identify the true underlying rank. (We omit the Horseshoe, Gamma, and IGG algorithm from this plot because their results nearly entirely overlap the Horseshoe+ results).

By examining Figure 1 the singular values resulting from the Horseshoe+ algorithm, we can easily identify the underlying rank-5 structure. However, we cannot identify this structure from examining the singular values of either the Gaussian or softImpute results.

\begin{figure}[H]
    \centering
    \makebox[1.2\textwidth][c]{\includegraphics[scale = .8]{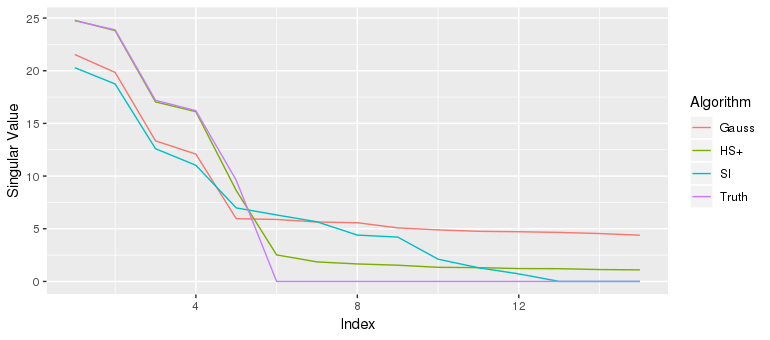}}
    \caption{Scree plot with the first 15 singular values of $\mathbf{\Theta} $ and $\mathbf{\hat{\Theta}}$ for three algorithms: Gaussian (Gauss), Horseshoe+ (HS+), and softImpute (SI).}
    \label{scree}
\end{figure}

\section{Matrix Completion Application: Board Game Ratings}

The preeminant website for hobby board game enthusiasts is \texttt{www.boardgamegeek.com}. The site catalogs information about every board game released, and has over one million registered users, who rate games on a continuous scale between 1 and 10. We compiled a data set of votes on the top 200 games by 5,000 randomly selected users who have rated least 5 of these games. (Users who had suspiciously rated over 200 games on the site were excluded). We represent this as a $5000 \times 200$ vote matrix with each row representing a user and each column representing a game.  

Here we compare the relative performance of various matrix completion algorithms on this real data. We randomly select 20\% of users and hold out one of each of their ratings to form a test set, $\mathcal{T}$. For the Gamma and softImpute algorithm, we use multiple values of $\beta$ and $\lambda$ respectively. For each algorithm, we calculate the standard error of its predictions as:  

$$\text{SE} = \sqrt{\sum_{\{i,j\} \in \mathcal{T}} ( \mathbf{Y}_{ij} - \mathbf{\hat{Y}}_{ij})^2}$$

where $\mathbf{Y}_{ij}$ player $i$'s actual rating on game $j$ and  $\mathbf{\hat{Y}}_{ij}$ is player $i$'s predicted rating on game $j$. 

We also compare the percentage of the total variation in rankings explained by the various models. We define this percentage as

$$\text{\% Explained} = 1 - \frac{\sum_{\{i,j\} \in \mathcal{T}} (\mathbf{Y}_{ij} - \mathbf{\hat{Y}}_{ij})^2}{\sum_{\{i,j\} \in \mathcal{T}} (\mathbf{Y}_{ij} - \mathbf{\bar{Y}})^2}$$
where $\mathbf{\bar{Y}}$ is the average rating across users and games in the test set. 

In order to predict user ratings, we include in each of these models row-wise intercepts representing users' levels of generosity, columnwise intercepts representing games' overall levels of quality, and and overall intercept. 

\begin{table}[H]
    \centering
\begin{tabular}{r l | c c }
 & Params & Test SE & \% Explained \\ \hline
Horseshoe+     & & 1.280 & 19.4\% \\ \hline
Horseshoe     & & 1.287 & 18.5 \% \\ \hline
IGG & $a,b,c = 1,.4,1$ & 1.283 & 19.0\% \\ \hline
\multirow{3}{*}{Gamma} & $\beta = 5$ & 1.293 & 17.8\%\\
 & $\beta = 30$ & 1.294 & 17.7\%\\
 & $\beta = 60$ & 1.293 & 17.8 \%\\ \hline
Gaussian & $V_0$ = 10 & 1.301 & 17.7 \% \\ \hline
\multirow{3}{*}{SoftImpute} & $\lambda = 5$ & 1.369 & 7.8 \%\\
 & $\lambda = 20$ & 1.287 & 18.5 \%\\
 & $\lambda = 40$ & 1.293 & 17.8 \%
\end{tabular}
    \caption{Chosen hyperparameters and test performance on board game rating data.}
    \label{boardgametable}
\end{table}

The relative performances of the various algorithms are largely consistent with the simulations, but in this case the global local priors perform only slightly better than softImpute, as long as we choose a suitable correct tuning parameter. It seems the greatest advantage here to the Horseshoe, Horseshoe+ and IGG formulations over softImpute are the lack of tuning parameters, indicating their convenience as default methods. The Gamma formulation seems to be more robust to its hyperparameter in this situation, but still worse performing overall.

As long as we are modeling this user-game vote matrix in terms of its decomposition into a de-facto user and game component matrices, we cannot resist conducting an exploratory factor analysis. This decomposition models user votes as depending on up to $K$ latent factors, which may be recognizable as game properties. After normalizing the columns of the component matrices $\mathbf{M}$ and $\mathbf{N}$, we can interpret each $\mathbf{M}_{ik}$ as the affinity of user $i$ for game property $k$, and we can interpret each $\mathbf{N}_{jk}$ as the degree to which game $j$ exhibits property $k$. 

When we position games according to their first two factor scores, $\mathbf{N}_{i1}$ and $\mathbf{N}_{i2}$ for $i \in \{1,\dots,200\},$ familiar patterns emerge.

We notice in Figure 2 that the coordinates seem to capture two aspects of games which are generally considered important by hobby gamers. Towards the bottom left are ``lighter'' games - they have shorter play time and easier to understand rules. To the top right of the horizontal axis are ``heavier'' games - these games appeal to ``core'' gamers who don't mind learning complex rule-sets and playing for many hours. Orthogonal to this direction, we find an axis that seems to correspond to a game's thematic content and flavor. To the bottom right, we find games with exciting and immersive themes and settings, and to the top left we find games with subtler and more traditionally European themes and settings.

The furthest game to the bottom left on the horizontal axis is Seiji Kanai's \textit{Love Letter}, a card game which only has a handful of different cards and only lasts a few minutes. \textit{Love Letter} has a ``Complexity'' rating on \texttt{boardgamegeek.com} of 1.20/5, an average of community votes. On the lower left, we also find Uwe Rosenberg's \textit{Bohnanza}, Marc Andr\'e's \textit{Splendor}, and Masao Suganuma's \textit{Machi Koro}, with complexity ratings of 1.67, 1.82, and 1.55 respectively. These games are all known for the light rule sets which make them suitable ``gateway games'' for newcomers to the hobby.

Meanwhile, to the top right we find notoriously formidable and lengthy games like Helge Ostertag's \textit{Terra Mystica}, Vlaada Chv\'atil's \textit{Through the Ages}, and Chad Jensen's \textit{Dominant Species}, with complexity ratings of 3.95, 4.17, and 4.03 respectively. The seeming outlier to the right of the image plot is Vlaada Chvatil's \textit{Mage Knight}, a game notorious for it's long playtime and complex movement and fighting mechanisms. \textit{Mage Knight} has a weight rating of 4.26 out of 5. 

Looking to the bottom right we find highly thematic games like Jonathan Gilmour and Isaac Vega's \textit{Dead of Winter}, a zombie survival game, Fantasy Flight's \textit{Star Wars Rebellion}, a game attached to an expensive intellectual property, and \textit{Mansions of Madness}, also by Fantasy Flight, one of many games set in a world of monsters inspired by H.P. Lovecraft. By contrast, at the top left are games like Andreas Seyfarth's \textit{Puerto Rico}, Uwe Rosenberg's \textit{Agricola}, and Bernd Brunnhofer's \textit{Saint Petersburg}. These games all take place in historical settings, in which players peacefully trade goods to maximize their economic returns. Hobby board game players sometimes call this style of game the ``Eurogame,'' as this type of economic resource-management game first become popular in Germany in the 1980s and 1990s, when Klaus Teuber's \textit{Settlers of Catan} helped bring about the modern renaissance in hobby games \citep{woods2012eurogames}. Various community members, depending on their preferences, are known either to heavily anticipate or sneer at ``yet another game about trading spices in the Mediterranean.'' 

We present these interpretations with due modesty considering that the total amount of variation in ratings explained by all of the factors combined is less than $20\%$ - it seems that either the noise in user ratings is much greater than the signal, or else the signal has a structure that cannot be fully captured by this bilinear model. Furthermore, the positions of the various games along the third principal component score axis and beyond elude our interpretation. 

\begin{figure}[H]
    \centering
    \makebox[1.2\textwidth][c]{\includegraphics[scale = 1.25]{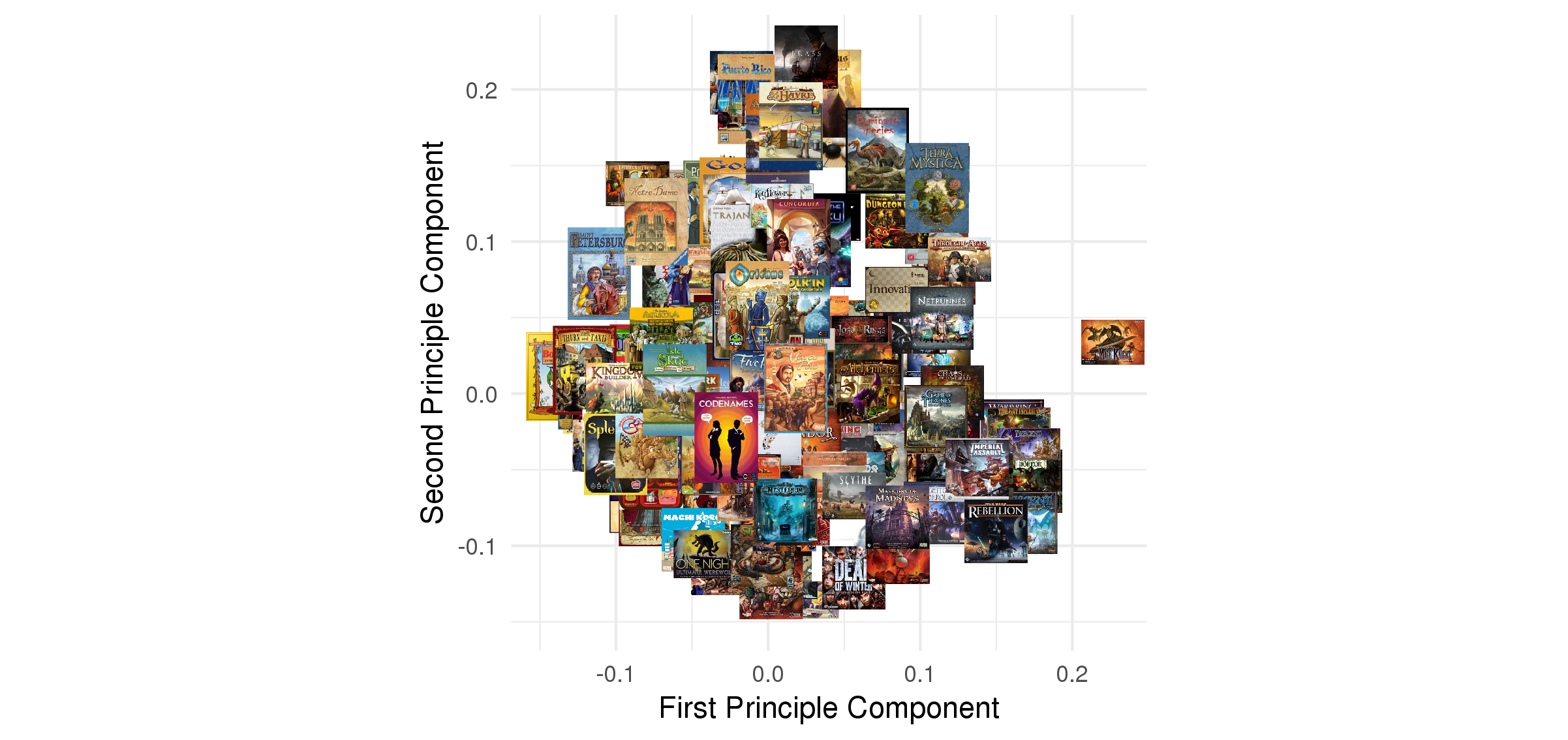}}
    \caption{Board games mapped according to their first two principal components using results from the Horsehoe+ algorithm.}
    \label{bgmap}
\end{figure}

\section{Tensor Completion Application: Color Image Recovery}

In this section, we test out the tensor completion functionality of our proposed algorithms. We use the representation of color images as $l \times w \times 3$ tensors; $l \times w$ is the size in pixels of the two-dimensional image, and there is one matrix slice for each of the red, green and blue color channels. By applying the low-rank tensor completion algorithm with intercepts for each dimension we can reasonably reconstruct the images even while most of the pixels are missing. 

In Table \ref{imagecomp}, we experiment with five different images (all covers of various classic board games) with different percentages of pixels removed uniformly at random. For images with few missing pixels, the output for each algorithm is visually indistinguishable. However, when a high percentage of pixels are missing, the Gaussian algorithm performs poorly, and the Gamma algorithm depends heavily on the choice of hyperparameter. This highlights the advantage of the global-local algorithms as out-of-the-box solutions.

In Table 7, we break down the outputs for one particular image (the cover of the board game \textit{Acquire}) into low rank reconstructions. We find the r-rank representations of the completed tensors by computing the 3-mode tensor Single Value Decomposition (SVD) of the form

$$\mathbf{\Theta} = \mathbf{U}\mathbf{S}\mathbf{V}^\top,$$
where $\mathbf{U}$ and $\mathbf{V}$ are orthogonal 3-mode tensors, and $\mathbf{S}$ is a 3-mode tensor whose matrix faces are diagonal matrices \citep{kilmer2013third,li2018package}. We then set the $(i,i,i)$th entry of $\mathbf{S}$ to 0 when $i > r$.

From this breakdown we can see that an accurate reconstruction requires information from a large number of principal components - in this sense the images are themselves not particularly low-rank. Thus we may not expect substantial gains in performance from sparse methods.

\begin{sidewaystable}
\label{imagecomp}
\newcommand{\myimagescale}{.45}
\begin{tabular}{c c | c c c c c c}
Original & Missing & Horseshoe+ & Horseshoe & IGG & \multicolumn{2}{c}{Gamma} & Gaussian \\
& & & & & $\beta = 5$ & $\beta = 30$ & $V_0 = 1$ \\
\includegraphics[scale = \myimagescale]{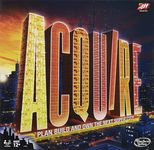} &
\includegraphics[scale = \myimagescale]{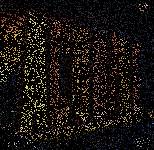} &
\includegraphics[scale = \myimagescale]{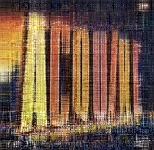} & 
\includegraphics[scale = \myimagescale]{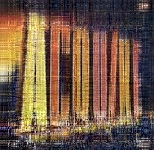} &
\includegraphics[scale = \myimagescale]{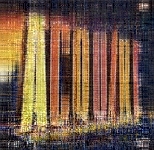} &
\includegraphics[scale = \myimagescale]{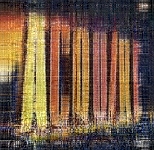} &
\includegraphics[scale = \myimagescale]{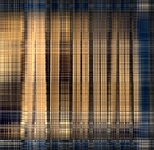} &
\includegraphics[scale = \myimagescale]{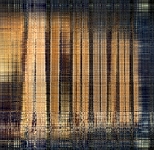} \\
& (80\% Missing) & & & & & & \\ 
\\
\includegraphics[scale = \myimagescale]{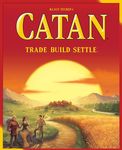} &
\includegraphics[scale = \myimagescale]{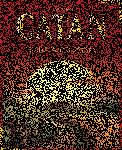} &
\includegraphics[scale = \myimagescale]{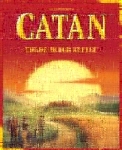} & 
\includegraphics[scale = \myimagescale]{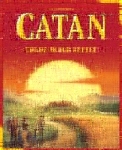} &
\includegraphics[scale = \myimagescale]{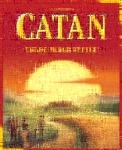} &
\includegraphics[scale = \myimagescale]{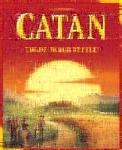} &
\includegraphics[scale = \myimagescale]{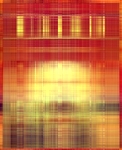} &
\includegraphics[scale = \myimagescale]{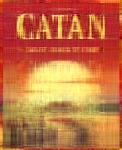} \\
& (60\% Missing) & & & & & & \\
\\
\includegraphics[scale = \myimagescale]{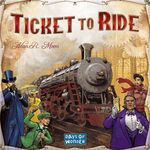} &
\includegraphics[scale = \myimagescale]{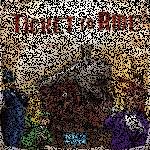} &
\includegraphics[scale = \myimagescale]{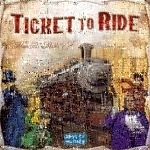} & 
\includegraphics[scale = \myimagescale]{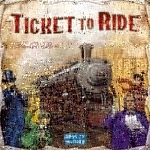} &
\includegraphics[scale = \myimagescale]{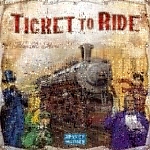} &
\includegraphics[scale = \myimagescale]{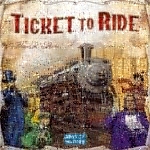} &
\includegraphics[scale = \myimagescale]{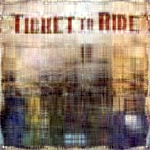} &
\includegraphics[scale = \myimagescale]{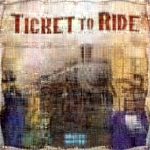} \\
& (40\% Missing) & & & & & & \\
\\
\includegraphics[scale = \myimagescale]{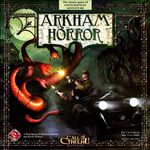} &
\includegraphics[scale = \myimagescale]{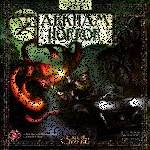} &
\includegraphics[scale = \myimagescale]{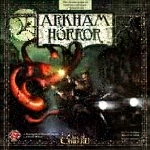} & 
\includegraphics[scale = \myimagescale]{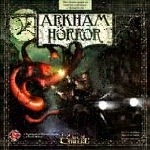} &
\includegraphics[scale = \myimagescale]{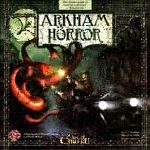} &
\includegraphics[scale = \myimagescale]{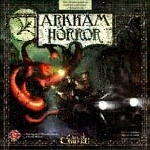} &
\includegraphics[scale = \myimagescale]{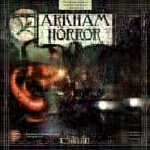} &
\includegraphics[scale = \myimagescale]{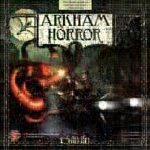} \\
& (20\% Missing) & & & & & & \\
\\
\includegraphics[scale = \myimagescale]{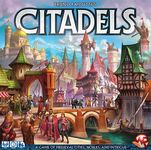} &
\includegraphics[scale = \myimagescale]{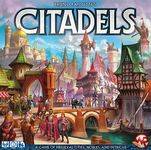} &
\includegraphics[scale = \myimagescale]{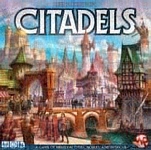} & 
\includegraphics[scale = \myimagescale]{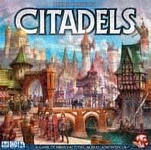} &
\includegraphics[scale = \myimagescale]{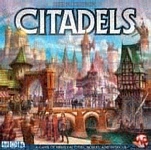} &
\includegraphics[scale = \myimagescale]{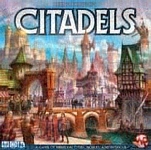} &
\includegraphics[scale = \myimagescale]{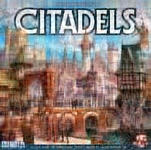} &
\includegraphics[scale = \myimagescale]{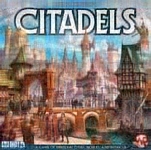} \\
& (0\% Missing) & & & & & & \\
\end{tabular}
\caption{Completions of five images with different levels of missing information by various Bayesian algorithms.}
\end{sidewaystable}

\begin{sidewaystable}
\label{dimensions}
\newcommand{\myimagescaletwo}{.45}
\begin{tabular}{m{2cm} | c c c c c c c}
& \multicolumn{7}{c}{Principal Components} \\
\\
 & 1 & 2 & 3 & 5 & 10 & 25 & 50\\ \hline
\\
\hfill Horseshoe+ & 
\includegraphics[scale = \myimagescaletwo]{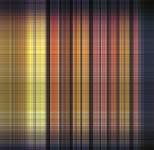} &
\includegraphics[scale = \myimagescaletwo]{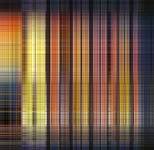} &
\includegraphics[scale = \myimagescaletwo]{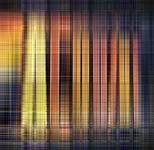} & 
\includegraphics[scale = \myimagescaletwo]{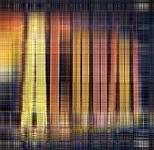} &
\includegraphics[scale = \myimagescaletwo]{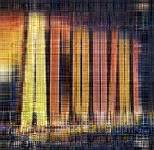} & 
\includegraphics[scale = \myimagescaletwo]{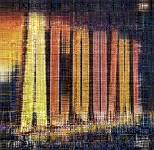} & 
\includegraphics[scale = \myimagescaletwo]{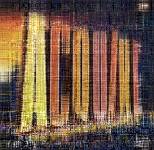} \\
\hfill Horseshoe & 
\includegraphics[scale = \myimagescaletwo]{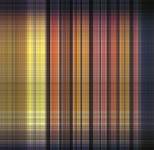} &
\includegraphics[scale = \myimagescaletwo]{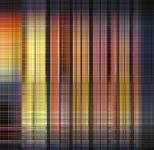} &
\includegraphics[scale = \myimagescaletwo]{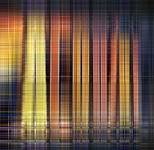} & 
\includegraphics[scale = \myimagescaletwo]{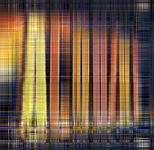} &
\includegraphics[scale = \myimagescaletwo]{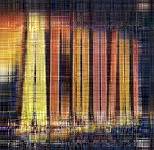} & 
\includegraphics[scale = \myimagescaletwo]{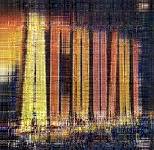} & 
\includegraphics[scale = \myimagescaletwo]{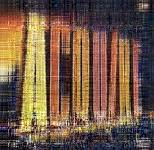} \\
\hfill IGG  & 
\includegraphics[scale = \myimagescaletwo]{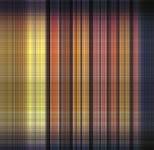} &
\includegraphics[scale = \myimagescaletwo]{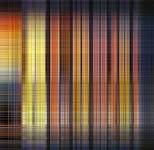} &
\includegraphics[scale = \myimagescaletwo]{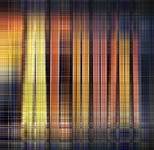} & 
\includegraphics[scale = \myimagescaletwo]{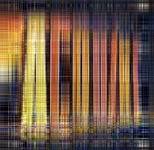} &
\includegraphics[scale = \myimagescaletwo]{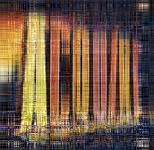} & 
\includegraphics[scale = \myimagescaletwo]{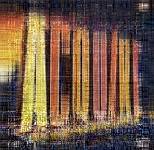} & 
\includegraphics[scale = \myimagescaletwo]{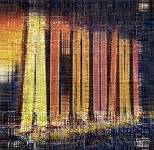} \\
\hfill Gamma &
\includegraphics[scale = \myimagescaletwo]{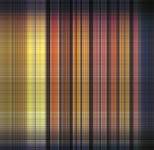} &
\includegraphics[scale = \myimagescaletwo]{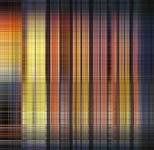} &
\includegraphics[scale = \myimagescaletwo]{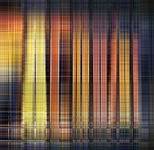} & 
\includegraphics[scale = \myimagescaletwo]{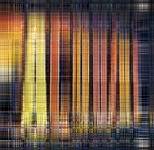} &
\includegraphics[scale = \myimagescaletwo]{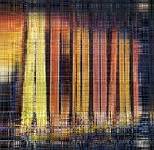} & 
\includegraphics[scale = \myimagescaletwo]{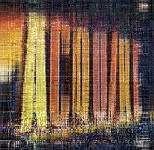} & 
\includegraphics[scale = \myimagescaletwo]{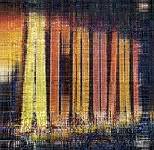}\\
\hfill Gaussian &
\includegraphics[scale = \myimagescaletwo]{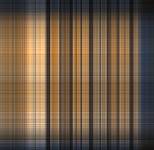} &
\includegraphics[scale = \myimagescaletwo]{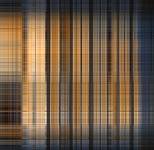} &
\includegraphics[scale = \myimagescaletwo]{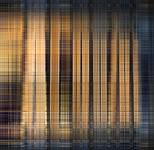} & 
\includegraphics[scale = \myimagescaletwo]{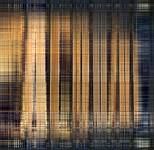} &
\includegraphics[scale = \myimagescaletwo]{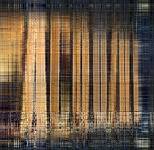} & 
\includegraphics[scale = \myimagescaletwo]{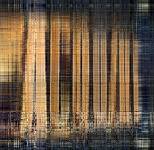} & 
\includegraphics[scale = \myimagescaletwo]{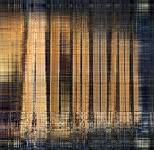}\\
\end{tabular}
\caption{Low-rank reconstructions of the completed image output from various Bayesian algorithms. }
\end{sidewaystable}

\section{Conclusion}

In this paper, we have presented three new prior formulations for modelling the components of matrix and tensor decompositions. These models outperform previous algorithms in the task of matrix and tensor completion in simulations and in real data examples. In addition, these algorithms require no tuning parameters, and thus provide convenient out-of-the-box solutions for matrix and tensor completion which automatically adapt to the underlying level of rank-sparsity in the data.

Beyond collaborative filtering and image completion, these factorization techniques can be directly applied to modify existing missing data imputation techniques based on matrix factorization \citep{sengupta2017}. The framework in this paper assumes there is no systematic relationship between missingness and response, and thus is directly applicable in the Missing Completely at Random setting \citep{rubin2004multiple}. Therefore, it would be interesting to extend these algorithms to account for data Missing Not at Random, as in \citet{sportisse2018imputation}. 

Although in this paper we have focused on the completion of matrices and tensors, global-local priors also provide a flexible structure for automatic rank-detection in  low-rank matrix and tensor decomposition problems in general. By elaborating on the structure of the priors on $\boldsymbol{\gamma},$ we can also encourage sparsity in the rows of the component columns and even the individual entries; see \citet{zhang2018abacus} for an application of such an extension to the problem of compression and change-point detection in multivariate time series. 

Most importantly, the connection between Bayesian variable selection in linear models and Bayesian low-rank matrix decomposition may point in additional promising directions. For example, in contrast to ``one-group'' priors including the global-local priors discussed in this paper, ``two-group'' or ``spike and slab'' priors \citep{mitchell1988bayesian} may be considered for the singular values of the response matrix. In particular, Spike and slab priors using ``non-local'' slab-priors \citep{rossell2017nonlocal} have increased power in finding null parameters in regression models, and thus may be useful in nullifying singular values. A wealth of results in sparse Bayesian linear models may yield analogous successes for low-rank models models in general.

\newpage
\nocite{*}
\bibliographystyle{chicago}
\bibliography{references}

\end{document}